# A physical method for investigating defect chemistry in solid metal oxides

Christian Rodenbücher[1,*], Carsten Korte[1], Thorsten Schmitz-Kempen[2], Sebastian Bette[2], Kristof Szot[2,3]

[1] Institute of Energy and Climate Research (IEK-14), Forschungszentrum Jülich GmbH, 52425 Jülich, Germany
[2] aixACCT Systems GmbH, 52068 Aachen, Germany
[3] A. Chełkowski Institute of Physics, University of Silesia, 41-500 Chorzów, Poland
* c.rodenbuecher@fz-juelich.de

**ABSTRACT**
The investigation of the defect chemistry of solid oxides is of central importance for the understanding of redox processes. This can be performed by measuring conductivity as a function of the oxygen partial pressure, which is conventionally established by using buffer gas mixtures or oxygen pumps based on zirconia. However, this approach has some limitations, such as difficulty regulating oxygen partial pressure in some intermediate-pressure regions or the possibility of influencing the redox process by gases that can also be incorporated into the oxide or react with the surface via heterogeneous catalysis. Herein, we present an alternative physical method in which the oxygen partial pressure is controlled by dosing pure oxygen inside an ultra-high vacuum chamber. To monitor the conductivity of the oxide under investigation, we employ a dedicated four-probe measurement system that relies on the application of a very small AC voltage, in combination with lock-in data acquisition using highly sensitive electrometers, minimizing the electrochemical polarization or electro-reduction and degradation effects. By analyzing the model material $SrTiO_3$, we demonstrate that its characteristic redox behavior can be reproduced in good agreement with the theory when performing simultaneous electrical conductivity relaxation (ECR) and high-temperature equilibrium conductivity (HTEC) measurements. We show that the use of pure oxygen allows for a direct analysis of the characteristic oxygen dose, which opens up various perspectives for a detailed analysis of the surface chemistry of redox processes.

## I. INTRODUCTION

Understanding the defect chemistry of solids is key to modeling and predicting their electrochemical properties. This statement is especially true for solid metal oxides, where the amount of point defects such as oxygen vacancies directly determine the electrochemical behavior. Hence, in recent decades, numerous comprehensive experimental and theoretical studies on the point defect chemistry of solid oxides have been conducted that have led to the development of a variety of technological applications that we nowadays use in daily life.[1,2] Examples can be found in the fields of gas sensors, high-temperature superconductors,[3] solid oxide fuel cells,[4] catalysts,[5] capacitors,[6] batteries[7] or neuromorphic computers.[8] Since the concentration of point defects in an oxide is directly related to the concentration of charge carriers, one of the most straightforward experimental methods to address the defect structure is to measure the electric conductivity as a function of temperature and oxygen activity.[9] Traditionally, this is realized by exposing an oxide to a buffer gas mixture,[10] such as $CO/CO_2$ or $H_2/H_2O$. Inert gases such as Ar or $N_2$ with a small amount of $O_2$ controlled by zirconia pumps[11,12] are also used. Although these methods allow for precise control of oxygen activity, direct information about the changes in the (non)-stoichiometry of the oxide is limited. Furthermore, there is the drawback that surface reactions could lead to uncontrolled redox reactions, e.g., for $KNbO_3$ it was found that the exposure to $CO_2$ induces more effective oxidation than the exposure to pure $O_2$.[13] It has also been found that $H_2$ can be incorporated into $SrTiO_3$, which could even reverse the effect of previously vacuum-induced reductions.[14]

Here, we present an alternate approach to the investigation of the defect chemistry of oxides using a physical method. The centerpiece of this approach is the replacement of the conventional gas mixtures by pure oxygen, which is dosed into a vacuum chamber. In this way, the oxygen activity can be directly controlled by varying the total oxygen pressure from atmospheric pressure down to high vacuum conditions, allowing for a detailed analysis of the redox reactions. Upon oxidation, the dose of the oxygen exposure to the sample can be determined, which is an important quantity for surface science and can be used to obtain information about the exchange reactions at the sample surface.

In order to demonstrate the capabilities of the described method, we present investigations of an SrTiO$_3$ single crystal. SrTiO$_3$ is especially suitable for this purpose, as it has become one of the most intensely investigated model materials for transition metal oxides with perovskite structures.[15–18] We present measurements of the resistance of a single crystal in four-probe geometry as a function of oxygen pressure at various temperatures in the range between 700 °C and 1000 °C. We show that the proposed method offers high reproducibility in conductivity measurements. In a proof of principle approach, we demonstrate that high-temperature equilibrium conductance (HTEC) and electrical conductivity relaxation (ECR) measurements can be performed simultaneously, the results of which are in good qualitative agreement with the expectations.

## II. THEORY

The macroscopic redox behavior of SrTiO$_3$ can be described by analyzing the equilibrium point defect concentration for a given temperature and oxygen partial pressure. This has been performed in detail, *e.g.*, by Moos and Härtl[19] and Denk et al.,[20] who elaborated on the defect chemistry of SrTiO$_3$ by carefully determining the equilibrium constants of the reactions between the point defects, as we will discuss briefly in the following. When SrTiO$_3$ is exposed to a reducing environment at moderate temperatures, twofold positively-ionized oxygen vacancies $V_O^{\cdot\cdot}$ are the major point defects:

$$O_O^x \rightleftharpoons V_O^{\cdot\cdot} + 2e + \frac{1}{2}O_2 \quad (1)$$

These are charge-compensated by electrons, which thus act as intrinsic donors, which lead to the so-called self-doping effect upon reduction. Accordingly, twofold negatively-ionized strontium vacancies serve as intrinsic acceptors whose generation is associated with the segregation of SrO, *e.g.*, as surface islands or as embedded Ruddlesden-Popper phases:[21]

$$Sr_{Sr}^x + O_O^x \rightleftharpoons V_{Sr}'' + V_O^{\cdot\cdot} + SrO \quad (2)$$

As even nominally undoped SrTiO$_3$ single crystals contain a non-negligible amount of extrinsic acceptors, such as Al, Cu, Fe, K, Mg or Mn, their contribution to the total defect equilibrium by the generation of holes upon ionization must also be taken into account:[22]

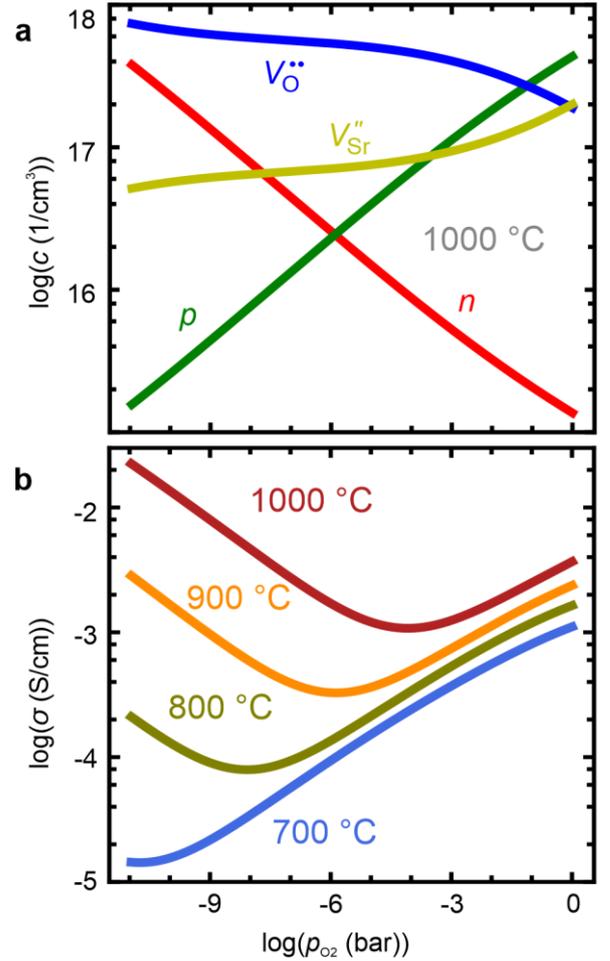

Figure 1. Point defect chemistry of SrTiO$_3$. a) Kröger-Vink diagram of defect concentrations calculated for nominally undoped SrTiO$_3$ at a temperature of 1000 °C; b) conductivities calculated from the concentration of electrons and holes and their mobilities for different temperatures.

$$A^x \rightleftharpoons A' + h \quad (3)$$

The concentration of thermally-generated electrons and holes in SrTiO$_3$ is very small due to a bandgap of 3.25 eV.[23] Hence, the equilibrium of generation and the recombination of electron-hole pairs is on the far left:

$$nil \rightleftharpoons e + h \quad (4)$$

By formulating the laws of mass actions for these four reactions together with the overall charge neutrality condition:

$$n + 2[V_{Sr}''] + [A'] = p + 2[V_O^{\cdot\cdot}], \quad (5)$$

where $n$ is the electron concentration and $p$ the hole concentration as well as the condition that the sum of neutral and ionized acceptors must be constant:

$$[A]=[A^x]+[A^{'}], \quad (6)$$

an equation system of six equations is generated that can be solved for a given temperature and oxygen activity level. This approach does not consider the presence of interstitials, singly ionized and neutral vacancies, which is regarded as being justified for sufficiently high temperatures.[24] In order to model the unintentional acceptor doping, we assumed an acceptor concentration of $10^{18}/cm^3$. Using the equilibrium constants derived by Moos and Härtl,[19] which are listed in the supplementary information, we calculated the defect concentration as a function of oxygen (partial) pressure using a computer-algebra software for numerical computing (Maplesoft, Waterloo, Canada). The resulting Kröger-Vink diagram is shown in Figure 1a for a temperature of 1000 °C. The diagrams for the temperatures between 700 °C and 900 °C can be found in the supplementary information.

The calculation predicts that the concentration of oxygen vacancies increases with decreasing oxygen pressure, and that the concentration of strontium vacancies decreases. Below a pressure of 0.8 bar, oxygen vacancies are the majority point defects. Regarding the concentration of electrons and holes, it can be seen that at atmospheric pressure, holes are dominant and that their concentration decreases with decreasing $O_2$ partial pressure, whereas the electron concentration increases. Hence, electrons become the dominant charge carriers below an $O_2$ partial pressure of $10^{-6}$ bar. In order to illustrate the consequences of the behavior of point defects on the electric characteristic of a solid oxide, we calculated the macroscopic conductivity $\sigma$ using the derived charge carrier concentration.[19] As the ionic transference number above 500 °C is small, ionic contributions are neglected:[25]

$$\sigma = en\mu_n + ep\mu_p \quad (7)$$

In Figure 1b, it can be seen that the conductivity initially decreases with decreasing oxygen pressure and then increases again after reaching a minimum. This intrinsic minimum marks the transition from a regime at high $O_2$ partial pressure, where p-conductivity prevails, to one at low $O_2$ partial pressure, where n-conductivity is dominant.[26] With decreasing temperature, the intrinsic minimum shifts to a lower $O_2$ partial pressure due to the relatively smaller concentration of electrons. As the conductivity is experimentally-observable, a direct comparison with the theory can be performed using an HTEC measurement.[27]

When the $O_2$ partial pressure in the surroundings of a solid oxide is rapidly changed, the material relaxes into a new equilibrium state. As this relaxation process directly relates to the oxygen incorporation, information about the surface exchange and chemical diffusion can be obtained by analyzing the conductivity traces, which are measured during the relaxation process using the ECR method.[28–31] The process of oxygen incorporation must be regarded as a two-step one comprising a surface reaction and the diffusion of $O^{2-}$ ions into the bulk.[15] It has been shown by Yasuda and Hishinuma[32] that the oxygen concentration during relaxation can be described by solving Fick's law in three dimensions:

$$\frac{\partial c}{\partial t} = D\nabla^2 c \quad (8)$$

where $c$ is the oxygen concentration in the oxide, which changes with time $t$ and $D$ is the chemical diffusion coefficient. Again, the model assumes an ideal crystal and does not account for inhomogeneous oxygen transport, which could occur, *e.g.*, along the grain boundaries or dislocations. In this model, the effect of the surface reaction is considered as boundary condition, following the approach of Crank[33] and introducing the surface exchange coefficient $k$. The oxygen concentration $c$ is directly related to the normalized conductivity:

$$\bar{\sigma}(t) = \frac{\sigma(t) - \sigma_0}{\sigma_\infty - \sigma_0}, \quad (9)$$

where $\sigma_0$ is the conductivity before the pressure step and $\sigma_\infty$ the conductivity in the new equilibrium measured immediately before the next pressure step was initiated. Hence, by performing least-squares fitting of the measured relaxation traces to the solution of Fick's law (see the supplementary information), the parameters $k$ and $D$ can be determined.

## III. EXPERIMENTAL SETUP

An aixDCA – Defect Chemistry Analyzer setup built by aixACCT Systems was used. It consisted of an ultra-high vacuum (UHV) chamber, with a quartz tube at one end, where the sample was positioned as shown in Figure 2a. The tube was surrounded by a temperature-controlled furnace, allowing for the performance of measurements of up to 1000 °C. Using a turbo pump in combination with an oil-sealed rotary pump serving as a backing pump, pressures down to $10^{-11}$ bar could be achieved. In order to perform measurements at different oxygen pressures, highly purified oxygen with a degree of 6.0 was continuously dosed into the vacuum chamber through a dosing valve (DV). The pressure was controlled using a Bayard-Alpert vacuum gauge[34] in the range of $10^{-11}$ bar to $10^{-6}$ bar ($P_{G1}$). The vacuum gauge was calibrated beforehand against a quadrupole mass spectrometer. As during measurement in this pressure regime, a constant flow of oxygen was present, and dedicated non-flammable oils had to be used in the rotary pump to avoid an explosion. Higher oxygen pressures of between $10^{-6}$ bar and 1 bar were achieved by closing the shut-off valve (SV) between the vacuum chamber and pump and filling in the oxygen until the desired pressure had been reached. Then, the dosing valve was closed and the measurements were performed in a static atmosphere. To reduce the oxygen pressure again, the shut-off valve was partially opened. This pressure control was performed by manually operating the valves. In the high-pressure regime, capacitance diaphragm gauges ($P_{G2}$-$P_{G3}$) were used to measure the pressure.

$SrTiO_3$ single crystals with a size of 10 mm x 4 mm x 1 mm (Shinkosha, Yokohama, Japan) were investigated in a four-probe geometry. Four electrodes were deposited on the top and backside of the crystal using a Pt paste. The distance between the inner electrodes was approximately 7 mm. Due to the preparation method of the electrodes based on applying the Pt paste by hand, a variation in the distance and shape resulting in a systematic error of the electrode dimensions of approximately 30 % had to be expected. Pt wires wrapped around the crystal were used as electrical connectors, as well as for the mechanical support of the sample, as is shown in Figure 2b. The electronic measurements were performed by applying an AC voltage at a frequency of 172.5 Hz provided by a generator to the outer electrodes. In order to avoid a potential stoichiometry polarization[35] or charging[36] during the measurement, a small amplitude of 4 mV was used. To measure the current, a current-voltage converter with an amplification of $10^5$ was employed. The electrical potentials $\Phi_0$, $\Phi_1$ and $\Phi_2$ at three electrodes were measured using sensitive electrometers connected via triaxially-screened cables, and one outer electrode was grounded (*cf.* Figure 2c). In this way, the voltage drops between the outer electrodes, inner electrodes, and between the outer and inner electrode on each side of the crystal, were recorded via a lock-in technique by a central processing unit (CPU). The total resistance of the sample, the bulk resistance, and the interface resistances were calculated using the RMS values of the potentials and the current:

$$R_{total} = \frac{\Phi_0}{I}, \qquad (10)$$

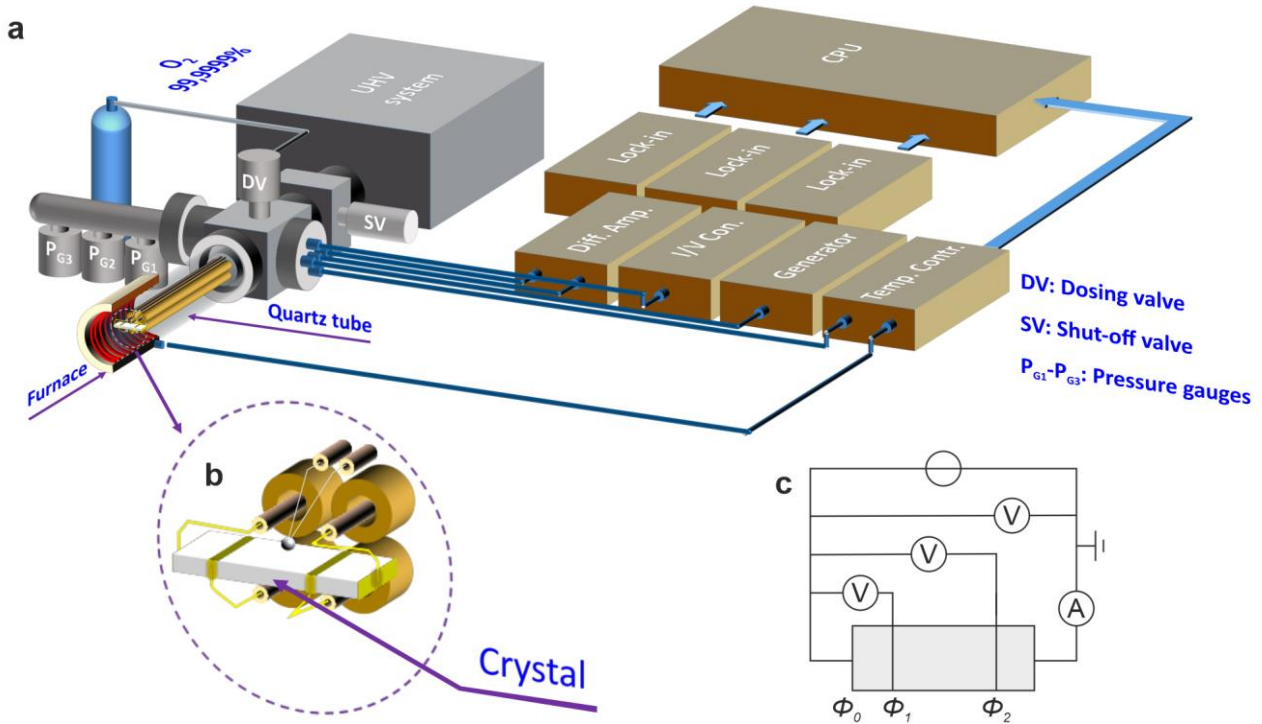

Figure 2. Illustration of the experimental setup: a) Schematics of the UHV chamber, where the crystal (b) is positioned and contacted in four-probe geometry; c) the measurement electronics and the schematic circuit diagram.

$$R_{bulk} = \frac{\Phi_1 - \Phi_2}{I}, \quad (11)$$

$$R_{interface\ 1} = \frac{\Phi_0 - \Phi_1}{I}, \quad (12)$$

$$R_{interface\ 2} = \frac{\Phi_2}{I}. \quad (13)$$

Subsequently, the macroscopic conductivity $\sigma$ was determined using the distance between the electrodes $d$ and the thickness $t$ and width $w$ of the sample:

$$\sigma = \frac{d}{R_{bulk} \cdot t \cdot w} \quad (14)$$

A detailed description and circuit simulation of the electronic measurement system can be found in the supplementary information.

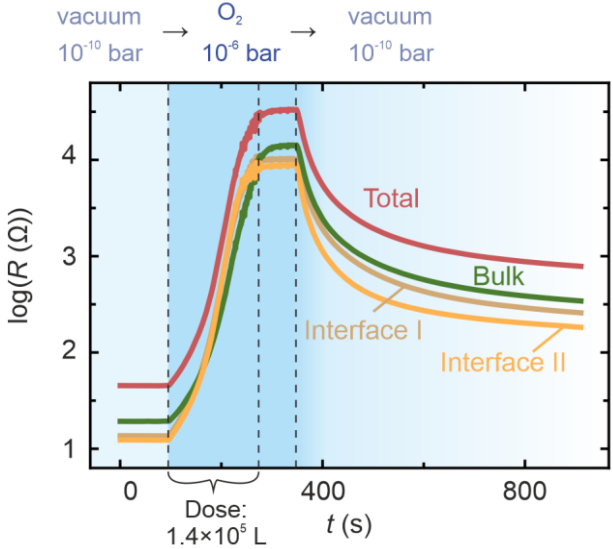

Figure 3. Resistance response of the SrTiO$_3$ crystal during exposure to 10$^{-6}$ bar oxygen for 255 s at 1000 °C.

## IV. RESULTS AND DISCUSSION

In order to illustrate the capacity of the measurement system, initially we present a response of the resistance of the sample on a pressure step. At a temperature of 1000 °C, the sample was equilibrated in a vacuum of $10^{-10}$ bar, *i.e.*, at maximum reducing conditions. Then, the oxygen pressure in the chamber was increased to $10^{-6}$ bar for 255 s. Subsequently, the chamber was pumped down to the maximum achievable vacuum again. The total resistance of the investigated SrTiO$_3$ crystal measured during this oxygen exposure is shown in Figure 3, together with the partial resistances of the bulk and the two interfaces. It can be seen that the resistance of the sample reacted very sensitively and increased immediately upon an increase in the oxygen pressure. After an oxidation time $t_{ox}$ of 190 s, the resistances reached a plateau and remained almost constant until the oxygen was pumped out of the chamber again. As the oxygen pressure during the oxidation was known, we could calculate the oxygen dose $D$ to which the sample was exposed until the new equilibrium was established:

$$D = p_{O_2} \cdot t_{ox} = 1.4 \times 10^5 \text{ L} \quad (15)$$

Comparing the resistances of the different sample regions, it can be seen that initially, the bulk resistance was higher than the resistances of the interfaces. During the exposure, this relationship was inverted and the increase in the interface resistances was faster than that of the bulk resistance. This shows that the surface of the crystal reacts faster than the bulk upon a change in oxygen activity.

Having demonstrated that we are able to measure the resistance changes in oxidation and reduction, we now focus on a detailed investigation of the defect chemistry across the entire accessible oxygen pressure range. We measured the resistance of the crystal while increasing the oxygen pressure in the chamber in a step-wise manner. Before starting the subsequent pressure step, we waited for the establishment of the new equilibrium. In this way, we could obtain ECR and HTEC information simultaneously. In Figure 4a, the resistance measurement as a function of time while increasing the oxygen pressure form $10^{-11}$ to $10^{-1}$ bar is shown for a temperature of 900 °C. It can be seen that all four resistances reacted similarly to the increasing oxygen pressure. The measured bulk resistance was then used to calculate the macroscopic normalized conductivity according to eqs. 9 and 14, as shown in Figure 4b. At low pressures, the equilibration time was in the range of several minutes and decreased significantly to values below 30 s at higher pressures. As is shown in detail in the supplementary information, the traces of normalized conductivity for each pressure step were simulated by fitting the solution of the diffusion equation in three dimensions after determining the starting time of each step using the tangent method. Figure 4c shows that the simulation agrees satisfactorily with the recorded conductivity data. Here, the pressure step from 1.1 mbar to 5 mbar is shown. The fit depends on two free parameters, namely the surface exchange coefficient $k$ and the diffusion coefficient $D$. This dependency is illustrated in the squared error matrix shown in Figure 4d. It can be seen that the minimum of the squared error $\chi$ is spread out over many orders of magnitude in $D$. Hence, the simulation is fairly insensitive with $D$ and only a lower limit of $10^{-8}$ m²/s can be given, whereas the simulation of $k$ is far more reliable. This effect is well known when analyzing ECR data and can potentially be improved in future by optimizing the sample geometry and experimental conditions.[37–39] Figure 4e shows the result of the simulation of $k$ at a temperature of 900 °C as a function of the oxygen pressure. In the low pressure regime, $k$ increases with increasing pressure before staying at the same order of magnitude above a pressure of $10^{-5}$ bar. Here, it should be kept in mind that the oxygen pressure was established by dosing pure oxygen inside the vacuum chamber. Hence, at low pressure, the mean free path of the oxygen molecules is very large and the decreased surface exchange coefficient may relate to the fact that the transport of oxygen from the vacuum atmosphere to the sample surface slowed down. By analyzing the ECR data of different temperatures between 700 °C and 1000 °C (see the supplementary information), an Arrhenius diagram of $k$ was obtained, as is shown in Figure 4f. In order to exclude the low pressure data, the mean of $k$ was calculated, taking only values above $10^{-4}$ bar for each temperature. Using linear regression, an activation energy of the surface exchange reaction of 0.23(4) eV was calculated, which is significantly lower than that obtained by optical studies.[31]

In order to extract the HTEC information from the recorded data, the macroscopic conductivity values were analyzed after reaching an equilibrium for each pressure step. They are plotted as a Brouwer diagram in Figure 5a. The data is in good qualitative agreement with the theoretical calculation in Figure 1b and with previous HTEC measurements, which used gas mixtures to establish the oxygen partial pressure. As expected, the $\log(\sigma)/\log(p_{O2})$ plot follows a slope of +¼ in the high-pressure regime. In the low-pressure one, it tends to approach a slope of −¼, but not enough data points could be obtained to derive clear evidence of this. By using the conductivity in the intrinsic minimum, where the contribution of holes and electrons to the conductivity is equal, the formation enthalpy of an electron-hole pair and, subsequently, the band gap energy $E_g^0$ at $T = 0$ K, can be estimated through an Arrhenius fit,[19] as is shown in Figure 5b (cf. Eq. 2 in the supplementary information). Here, we exclude the measurement performed at 700 °C, as the conductivity minimum could not be clearly identified because it was too close to the lower limit of the achievable $O_2$ pressure range. In order to calculate the band gap energy from the slope of the fit, the activation energy of the product of the density of the states in the valence band and the hole mobility was used as calculated by Moos and Härtl,[19] and the ionic partial conductivities were neglected. We obtain a value of 2.50(2) eV, which is lower than the literature value of the band gap of 3.25 eV.[23] This difference could relate to the influence of ionic conductivity in the intrinsic minimum. *E.g.* Ohly et al.[40] extrapolated the linear fit of the Brouwer diagram in the n and p region to determine the conductivity in the minimum. We did not compensate for this contribution because a typical plateau close to the intrinsic minimum was not observed. It should also be noted that the band gap at the surface can be much smaller than that of the bulk. By X-ray photoelectron microscopy (XPS), we were able to find a value of approximately 2.7 eV (see the supplementary information). As it has been found that the surface layer of $SrTiO_3$ plays a dominant role in electrical transport,[41] the deviation in the estimated band gap from the bulk value may reflect that the macroscopic conductivity is primarily determined by the conductivity of the surface layer.

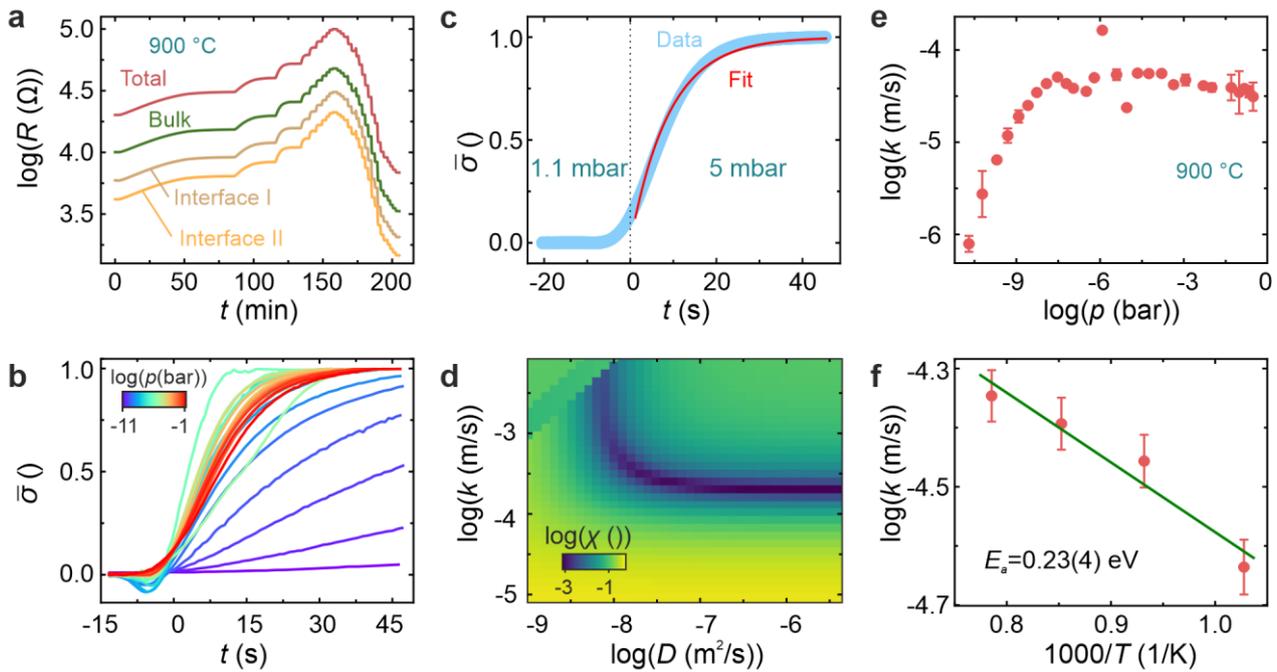

Figure 4. Electronic conductivity relaxation analysis at 900 °C. a) Resistance of the sample upon stepwise exposure to oxygen from $10^{-11}$ to 0.3 bar; b) normalized conductivity calculated from the "Bulk" data in (a); c) comparison between the data and fit shown for the oxygen step from 1.1 to 5 mbar; d) squared error matrix of the diffusion coefficient $D$ and the surface exchange coefficient $k$ used for generating the fit in (c); e) estimated surface exchange coefficient as a function of the oxygen pressure; f) Arrhenius plot of the surface exchange coefficient in the high-pressure regime ($p>10^{-4}$ bar).

Similarly, the activation energy in the oxidizing regime was calculated by performing an Arrhenius fit using the conductivity measured at $10^{-1}$ bar. As is shown in Figure 5b, an activation energy of $E_a^{ox}$ = 0.42(1) eV is obtained, resulting in an oxidation enthalpy of $\Delta H^{ox} = 2E_a^{ox}$ = 0.84(2) eV, which is also slightly lower than in the literature.[19,42] A direct comparison of the Brouwer diagram obtained in this study with literature data can be found in the supplementary information, which shows that our method reproduces the pressure dependence of the conductivity with comparable data quality. However, a significant variation between the data published in different papers concerning the magnitude of the conductivity, as well as the slopes of the pressure-dependent conductivity curves and the position of the intrinsic minimum is also present.[11,12,19,43,44] Despite these discrepancies, which may be related to a variation in crystal production and preparation[45,46] and could be systematically investigated in future, our proof-of-concept approach shows that the characteristic Brouwer diagram of $SrTiO_3$ can be reproduced using pure oxygen without the need of buffering gas mixtures.

Although we find apparent good agreement between our macroscopic measurement and the theory of point defect chemistry, we would like to note a few caveats relating to the conductivity of $SrTiO_3$, whose detailed investigation lays far beyond the scope of the present paper. Using techniques with high spatial resolution such as local-conductivity atomic force microscopy (LC-AFM), it has been shown that electronic transport in $SrTiO_3$ is highly inhomogeneous on the nanoscale and that dislocations play a decisive role as easy reduction sites.[46–48] By thermal reduction, it is even possible to induce an insulator-to-metal transition in $SrTiO_3$, *i.e.*, generating a conduction state that lies outside of the Brouwer diagram.[49] These effects may be of particular relevance when building nanoscale devices, such as for resistive switching applications.[50,51]

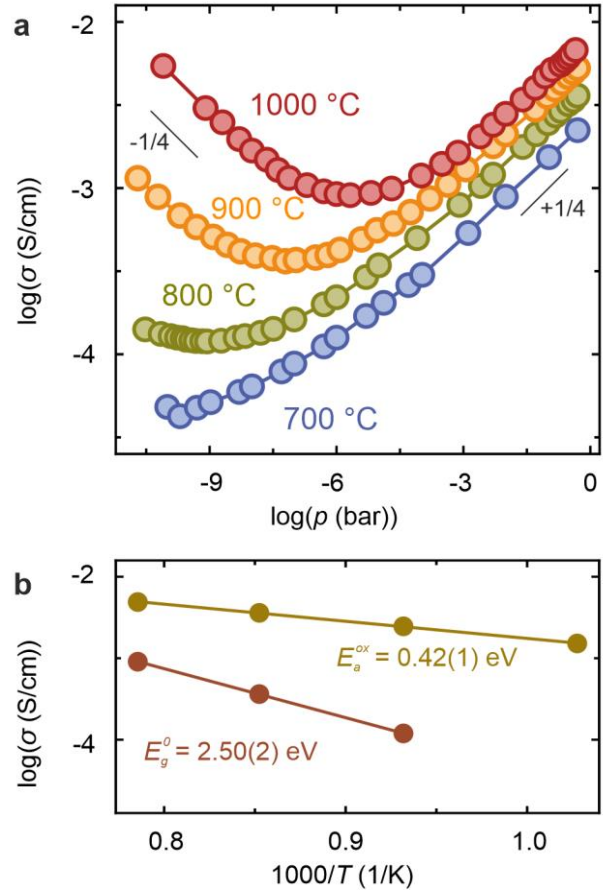

Figure 5. HTEC analysis of the $SrTiO_3$ crystal at different temperatures. a) Brouwer diagram of the conductivity as function of partial pressure. b) Arrhenius plot of the conductivity in the extrinsic minimum and in the high-pressure regime.

The possibility of analyzing the point defect chemistry diagrams through the controlled exposure of the clean $SrTiO_3$ surface at precisely-defined oxygen doses presented here opens up a new path in the search for a "common denominator" between data obtained through macroscopic electrical characterization and that obtained using surface sensitive experimental methods and theoretical *ab initio* calculations. The local stoichiometry, chemical composition, crystallographic structure (in-plane), electronic structure, local electrical transport phenomena and the the homogeneity of the redox reaction can be analyzed in detail for the definition of the boundary conditions for the reaction surface/gas phase. For instance, it can be shown by XPS studies of the $SrTiO_3$ crystal after different reduction and reoxidation steps that the electronic structure of the surface is significantly changed. Upon thermal reduction, the valences of the Ti ions are partially changed from +4 to +3 and +2 and occupied states in the band gap evolve, and also a portion of the oxygen ions can be removed from the crystal.[46] These changes are reversed by reoxidation, showing that the oxygen vacancies are refilled and the charge distribution returns to the initial state (see the supplementary information). Combining these data on the electronic structure with other surface-sensitive techniques such as LC-AFM, low-energy electron diffraction (LEED) or scanning tunneling microscopy (STM), providing information about the atomic arrangement of the surface, models regarding the oxygen exchange reactions can be directly verified and correlated with measurements of macroscopic conductivity.

## V. CONCLUSIONS

In summary, in this study we have shown that the point defect chemistry of solid metal oxides can be investigated using a typical surface physics method, which relies on exposing the sample to controlled doses of pure $O_2$ in a vacuum chamber. This method eliminates potential catalytic surface reactions or a significant proton incorporation, which can occur when gas mixtures are used. By measuring the conductivity of $SrTiO_3$ as a function of time and oxygen pressure, the characteristic behavior, as modeled by point defect chemistry, could be reproduced, showing that the method is fully capable of investigating the redox behavior of solid oxides. Using a four-probe measurement technique, in combination with an AC lock-in technique, detailed information about the behavior of the surface region and bulk can be obtained. Although the interface between the Pt electrodes and $SrTiO_3$ crystal shows a different kinetic response than the bulk of the sample, their pressure-dependent defect behavior is the same. Due to the use of pure oxygen, the introduced setup provides extensive information about the oxygen dose, which can help illuminate the surface chemistry, especially concerning the kinetics of the exchange reactions for the bulk and interfaces. As the physical method of investigating defect chemistry introduced here is compatible with other measurement techniques that require UHV conditions such as mass spectrometry, photoelectron spectroscopy or scanning probe methods, we expect that our approach can help bridge the gap between macro- and nanoscale understanding of electronic transport in transition metal oxides by combining conductivity measurements with *in situ* analytics in the future.

## SUPPLEMENTARY MATERIAL

See the supplementary material for details on the theory of point defect chemistry, the electronic measurement system, data processing, ECR simulation and background information regarding the analysis of $SrTiO_3$ single crystals using surface-sensitive techniques.

## ACKNOWLEDGMENTS

We gratefully acknowledge C. Wood for proofreading the manuscript.

## DATA AVAILABILITY

The data that support the findings of this study are available from the corresponding author upon reasonable request.

Supplementary information to:
**A physical method for investigating defect chemistry in solid metal oxides**

      I.    Calculation of defect concentrations

Following the approach of Moos and Härdtl [1], we calculated the point defect concentrations for SrTiO$_3$ at different temperatures. The calculation assumes a homogeneous material with statistically-distributed, non-interacting point defects (ideal solution). As doubly-ionized vacancies can be expected to be the dominant point defects in the investigated temperature range, we did not take singly-ionized or neutral oxygen and strontium vacancies into account.

Under these assumptions, the point defect chemistry of acceptor-doped SrTiO$_3$ is determined by the following equations, derived from the law of mass actions, the electroneutrality condition and the conservation of acceptor concentration:

$$[V_O^{\cdot\cdot}]n^2 p_{O_2}^{1/2} = K_{red}^0 \exp\left(-\frac{\Delta H_{red}}{kT}\right) \quad (1)$$

$$np = N_C(T)N_V(T)\exp\left(-\frac{E_g(0\text{ K}) - \beta_g T}{kT}\right) \quad (2)$$

$$\frac{[A']p}{[A^x]} = K_A^0 \exp\left(-\frac{E_A}{kT}\right) \quad (3)$$

$$[V_{Sr}''][V_O^{\cdot\cdot}] = K_S^0 \exp\left(-\frac{E_S}{kT}\right) \quad (4)$$

$$n + 2[V_{Sr}''] + [A'] = p + 2[V_O^{\cdot\cdot}] \quad (5)$$

$$[A] = [A'] + [A^x] \quad (6)$$

with

- $[V_O^{\cdot\cdot}]$: Concentration of two-fold ionized oxygen vacancies
- $[V_{Sr}'']$: Concentration of two-fold ionized strontium vacancies
- $[A]$: Total concentration of acceptors
- $[A']$: Concentration of singly-ionized acceptors
- $[A^x]$: Concentration of neutral acceptors
- $n$: Concentration of electrons
- $p$: Concentration of holes
- $p_{O_2}$: Oxygen partial pressure
- $k$: Boltzmann constant
- $T$: Temperature
- $K_{red}^0$: Frequency factor for the formation constant of doubly-ionized oxygen vacancies
- $\Delta H_{red}$: Reduction enthalpy
- $N_C(T)$: Effective density of states in the conduction band
- $N_V(T)$: Effective density of states in the valence band
- $E_g(0\text{ K})$: Band-gap energy
- $\beta_g$: Temperature coefficient of the band-gap
- $K_A^0$: Frequency factor of the formation constant of singly-ionized acceptors
- $E_A$: Activation energy of the formation constant of singly-ionized acceptors
- $K_S^0$: Frequency factor of the formation constant of doubly-ionized strontium vacancies
- $E_S$: Activation energy of the formation constant of doubly-ionized strontium vacancies

We used the following values, taken from [1]:

| | |
|---|---|
| $k$ | $8.617 \times 10^{-5}$ eV/K |
| $K_{red}^0$ | $5 \times 10^{71}$ bar$^{1/2}$/cm$^9$ |
| $\Delta H_{red}$ | 6.1 eV |
| $N_C(T)$ | $4.1 \times 10^{16}$ cm$^{-3}(T/\text{K})^{1.5}$ |
| $N_V(T)$ | $3.5 \times 10^{16}$ cm$^{-3}(T/\text{K})^{1.5}$ |
| $E_g(0\text{ K})$ | 3.17 eV |
| $\beta_g$ | $5.66 \times 10^{-4}$ eV/K |
| $K_A^0$ | $N_V(T)$ |
| $E_A$ | 0.94 eV |
| $K_S^0$ | $3 \times 10^{44}$/cm$^6$ |
| $E_S$ | 2.5 eV |

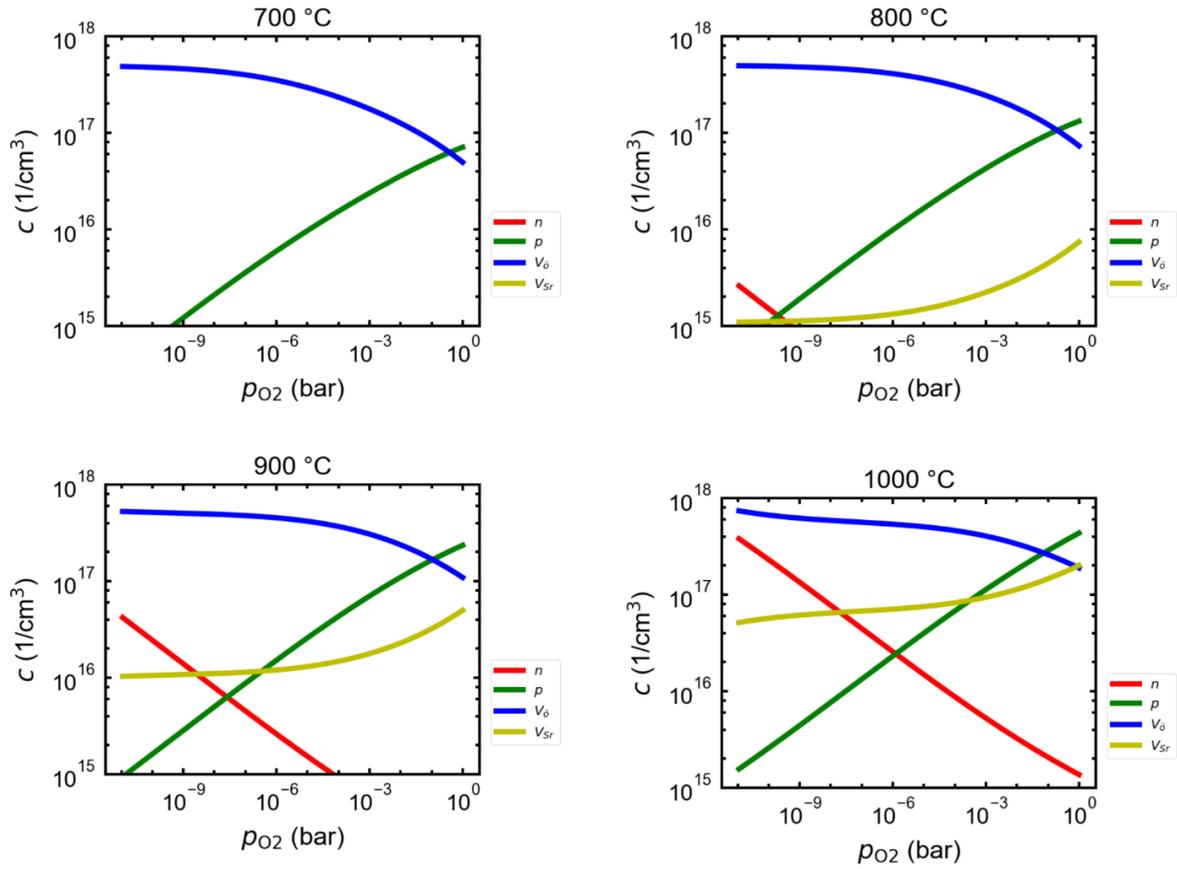

Figure S1. Concentration of the main defects in SrTiO$_3$ as calculated by the defect chemistry for different temperatures.

As nominally undoped SrTiO$_3$ single crystals typically contain a non-negligible amount of acceptor-type contaminants, we assumed an acceptor concentration of $[A] = 10^{18}$/cm$^3$. By solving the system of equations (1)-(6) for different temperatures and O$_2$ partial pressures, we calculated the defect concentrations, as shown in Figure S1. It can be seen for all of the investigated temperatures and O$_2$ partial pressures that oxygen vacancies are the major point defects. With increasing temperature, the concentration of strontium vacancies increases and at 1000 °C and under oxidizing conditions, their concentration is approximately equal to the concentration of oxygen vacancies. Furthermore, the concentration of electrons and holes increases with higher temperatures. As the electron concentration increases faster than the hole concentration, electrons become the dominant charge carries at low oxygen pressures. Hence, the transition between n-type and p-type conductivity is shifted to higher oxygen pressures with increasing temperature. This allows for the investigation of the intrinsic conductivity minimum in a pressure regime, which can be reached with the used vacuum apparatus. In the intrinsic conductivity minimum, the conductivity contributions of electrons and holes are equal, which allows for the band gap energy to be estimated.

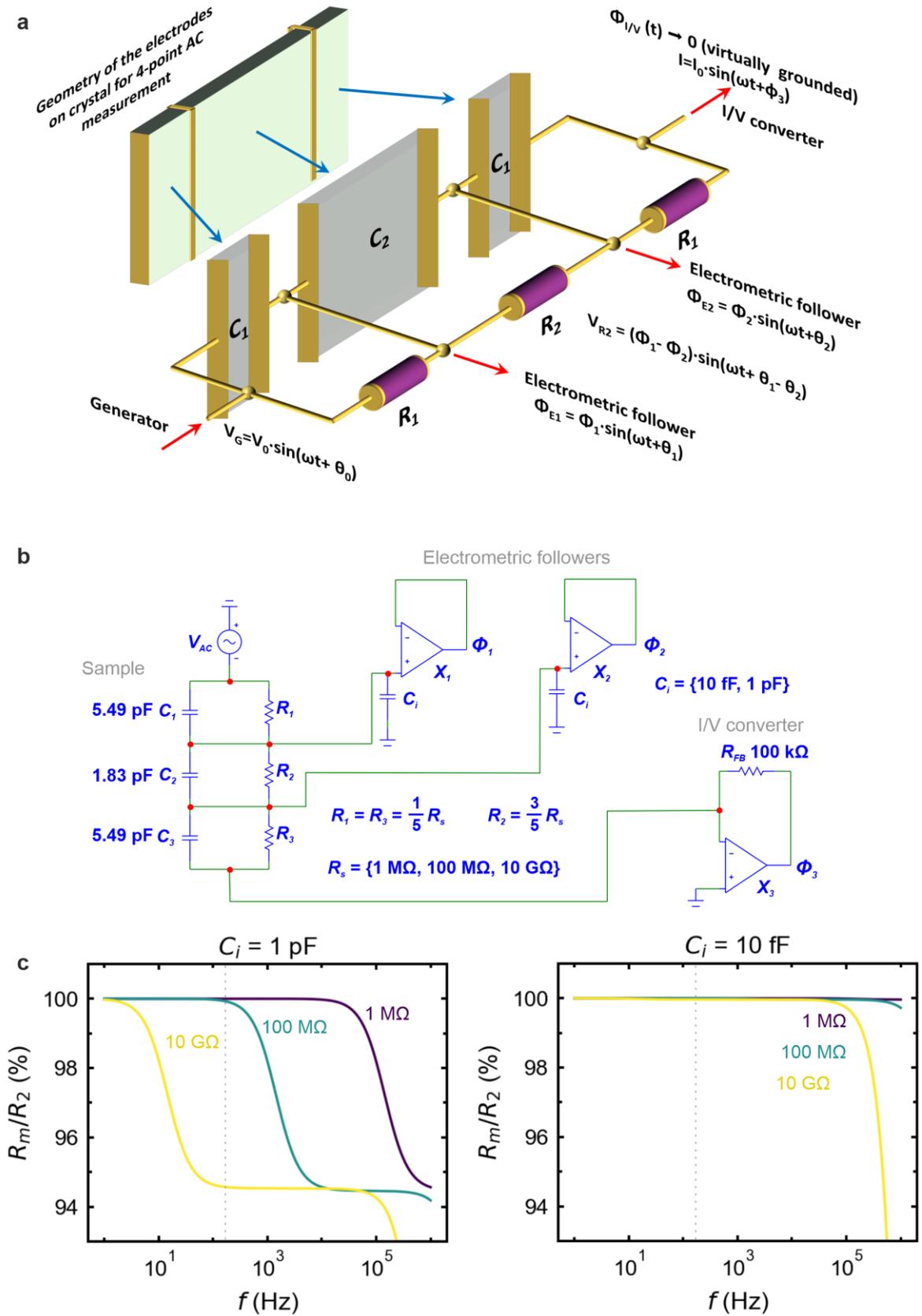

Figure S2. Measurement of the resistance using the AC method. a) Geometry of the sample and equivalent circuit; b) outline of the simulation circuit comprising the sample, electrometric followers and I/V converter; c) ratio between the measured resistance $R_m$ and the real resistance $R_2$ as calculated by the circuit simulation for different sample resistances and different input capacitances $C_i$ of the operational amplifiers.

## II. Principle of resistance measurement

To measure the macroscopic conductivity of the sample, we used a four-probe geometry, as is illustrated in Figure S2a. As we applied an AC voltage to determine the resistances by measuring the voltage drop between the electrodes, the resistive and capacitive effects of the sample must be taken into account when analyzing the electronic properties of the measurement system. To model the sample by means of electric circuit simulation, we assume that the sample is divided into three parts by the inner electrodes and simulated each part by an R-C element, as is shown in Figure S2a. We assume that the inner part of the sample is 3/5 of the total length and hence the outer parts of the sample have a length of 1/5 of the total length each. The total geometric capacitance of the sample can be estimated using width $w = 4$ mm, thickness $t = 1$ mm, length $l = 10$ mm, dielectric constant $\varepsilon_r = 310$ [2] and vacuum permittivity $\varepsilon_0 = 8.854 \times 10^{-12}$ F/m:

$$C_{total} = \varepsilon_0 \varepsilon_r \frac{w\,t}{l} = 1.098 \text{ pF}. \tag{7}$$

The capacitances of the three regions are

$$C_1 = C_3 = 5 \cdot C_{total} = 5.49 \text{ pF} \tag{8}$$

and

$$C_2 = \frac{5}{3} \cdot C_{total} = 1.83 \text{ pF}. \tag{9}$$

The resistances of the three part can be calculated analogously

$$R_1 = R_3 = \frac{1}{5} \cdot R_{total} \tag{10}$$

and

$$R_2 = \frac{3}{5} \cdot R_{total}. \tag{11}$$

We performed the simulation for three different total resistances $R_{total}$ of 1 MΩ, 100 MΩ and 1 GΩ, in order to illustrate the behavior of the measurement system.

The electronic circuit diagram of the measurement system is shown in Figure S2b. The potential at the inner electrodes $\Phi_1$ and $\Phi_2$ was measured using two electrometric followers and current by a transimpedance amplifier with a feedback resistance of $R_{FB} = 100$ kΩ, converting the current into a proportional voltage $\Phi_3$. Using a small AC voltage of 10 mV to measure the resistance of a highly resistive sample requires the use of electrometric followers with extremely low input capacitances $C_i$. To illustrate this effect, we simulated the potentials $\Phi_1$, $\Phi_2$ and $\Phi_3$ for two different input capacitances of 1 pF and 10 fF as function of the frequency of the AC voltage using the Microcap software. The measured bulk resistance $R_m$ was then calculated as:

$$R_m = \frac{Re(\Phi_2 - \Phi_1)}{Re(\Phi_3/-R_{FB})} \quad (12)$$

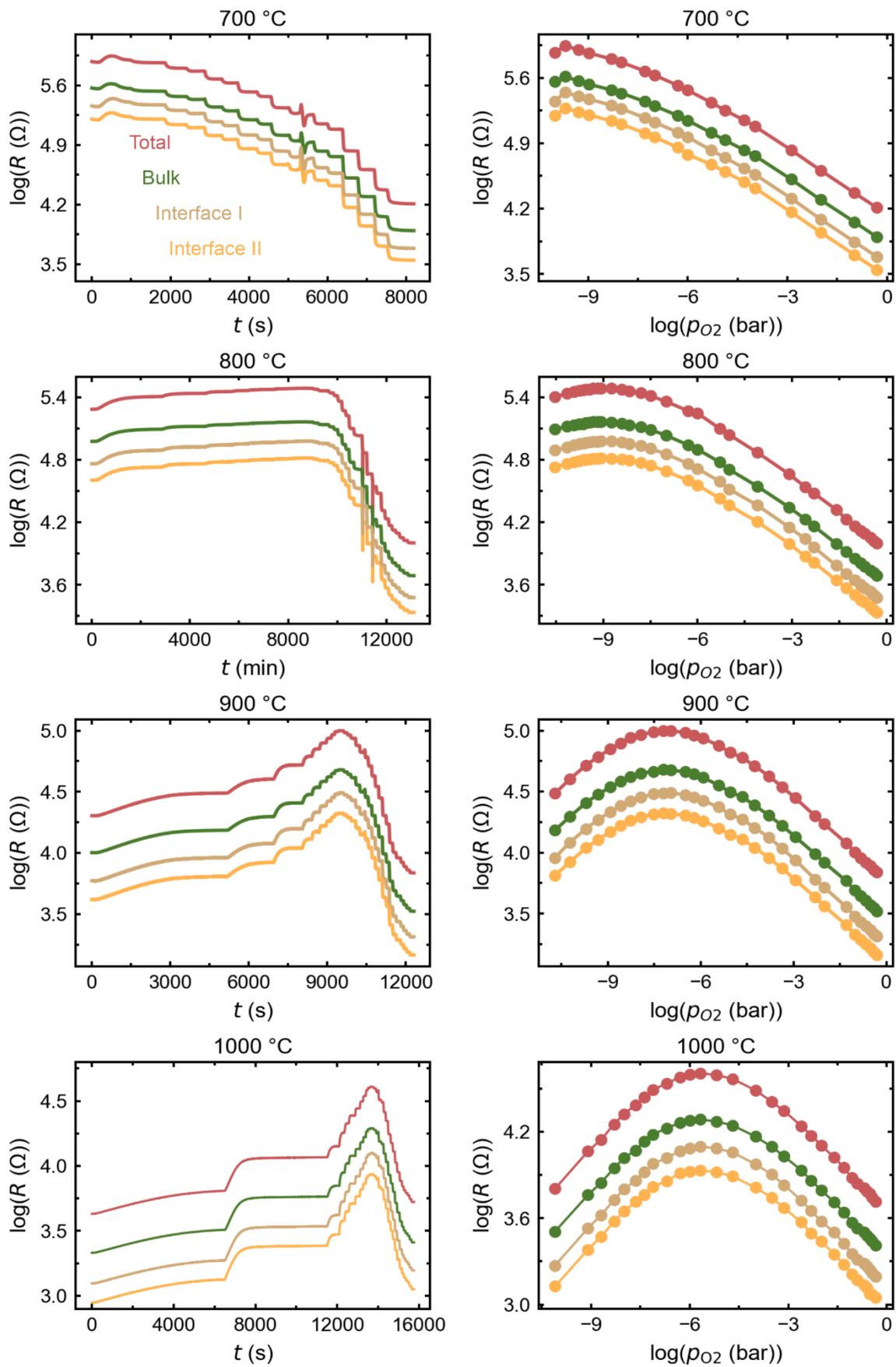

Figure S3. Resistance of the SrTiO$_3$ sample obtained while dosing oxygen in the vacuum chamber. The left column shows the raw data and the right column the equilibrium resistance as a function of the oxygen pressure.

In Figure S2c, the results of the simulation are shown as ratios between the (simulated) measured resistance and the real resistance of the inner part, $R_2$. It can be seen that significant deviations from the real resistance are present when assuming an input capacitance of the electrometric followers of $C_i = 1$ pF, which is a typical value of conventional operation amplifiers. With increasing sample resistance, the deviations become more pronounced, and for samples with a total resistance of more than 100 MΩ, deviations of several percent are present at the AC frequency of 173 Hz, which was used in the present study. In order to minimize this effect, we developed electrometric followers with a reduced input capacitance of $C_i < 10$ fF. The simulation shows that when using electrometric followers with $C_i = 10$ fF, the deviation between the real measured resistance is to be expected at less than 0.1% at frequencies of up to several 10 kHz, even when dealing with highly resistive samples.

### III. Extraction of the equilibrium resistance from the raw data

In this section, we illustrate how the conductivity data presented in the main text was derived from the raw data of the resistance measurements. The resistance of the SrTiO$_3$ sample was measured at four different temperatures, from 700 °C to 1000 °C, at different oxygen partial pressures. After stabilization of the temperature under vacuum conditions, oxygen was dosed in the chamber in steps. After each pressure step, we waited for the sample to reach equilibrium before initiating the following step. This was performed by continuously observing the resistance trace on the measurement computer. In Figure S3, the raw data of the total sample resistance are shown together with the resistance contributions of the bulk and the two interfaces. The left column shows the raw data as a function of the measurement time, illustrating the change in the resistance at each pressure step. The right column shows the resistance values taken when the equilibrium was reached as a function of the oxygen pressure, showing that the resistances exhibits the expected behavior, displaying a maximum of resistance as a consequence of the transition between n-type and p-type conductivity as discussed above. The fact that the resistance contributions from the bulk and interfaces react in the same manner on the oxygen pressure indicates that the sample was electrically-symmetrical during the measurements and no polarization or phase shift effects played a significant role. This shows that the method employed is well-suited for the electrical characterization of solid oxide samples. For the calculation of the macroscopic conductivity $\sigma$, as displayed in the main manuscript, the values of the bulk resistance $R_{bulk}$ were used.

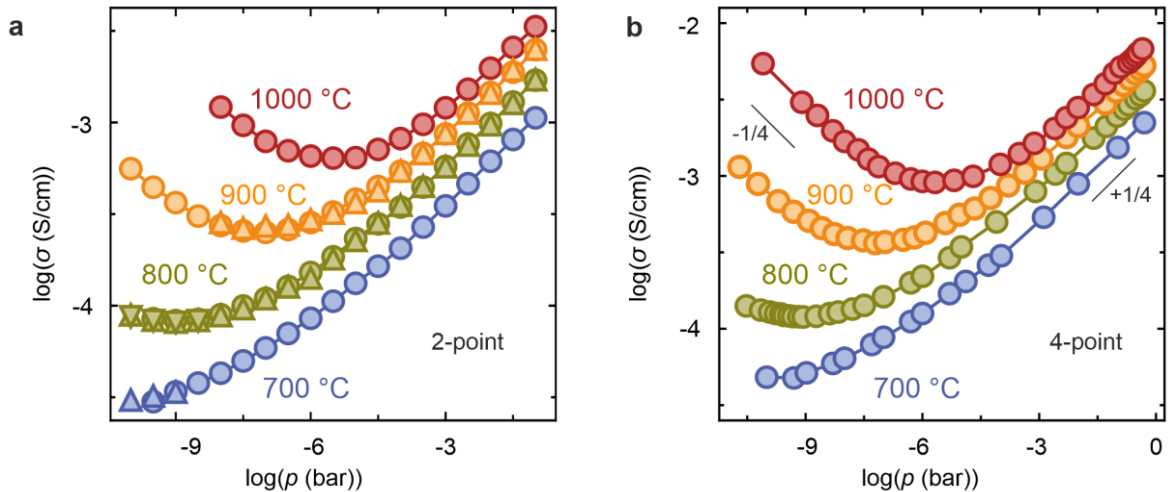

Figure S4. (a) Brouwer diagrams measured in 2-point geometry; and (b) 4-point geometry. The different symbols represent data obtained in different measurement runs.

## IV. Influence of measurement geometry

As was previously described, the HTEC data were measured using a very small AC voltage in conjunction with a lock-in technique to avoid the effects of stoichiometric polarization or electrodegradation, which could occur if large DC voltages were applied. To illustrate that this approach was successful, in Figure S4, we compare the Brouwer diagrams measured in a two-point geometry using only the outer electrodes on the sides of the sample and those measured in four-point geometry, as described in the main manuscript. It can be seen that the data from both geometries show a comparable trend and reproduce the characteristic pressure and temperature dependence of the conductivity. The conductivity measured in the 2-point geometry is slightly higher than that measured in the 4-point one, which may be caused by a small influence of the contact resistance at the electrode-oxide interface but also difficulties in electrode deposition, which was performed manually using Pt paste, and may have introduced a systematic error in the estimation of the electrode distance and subsequently of the conductivity.

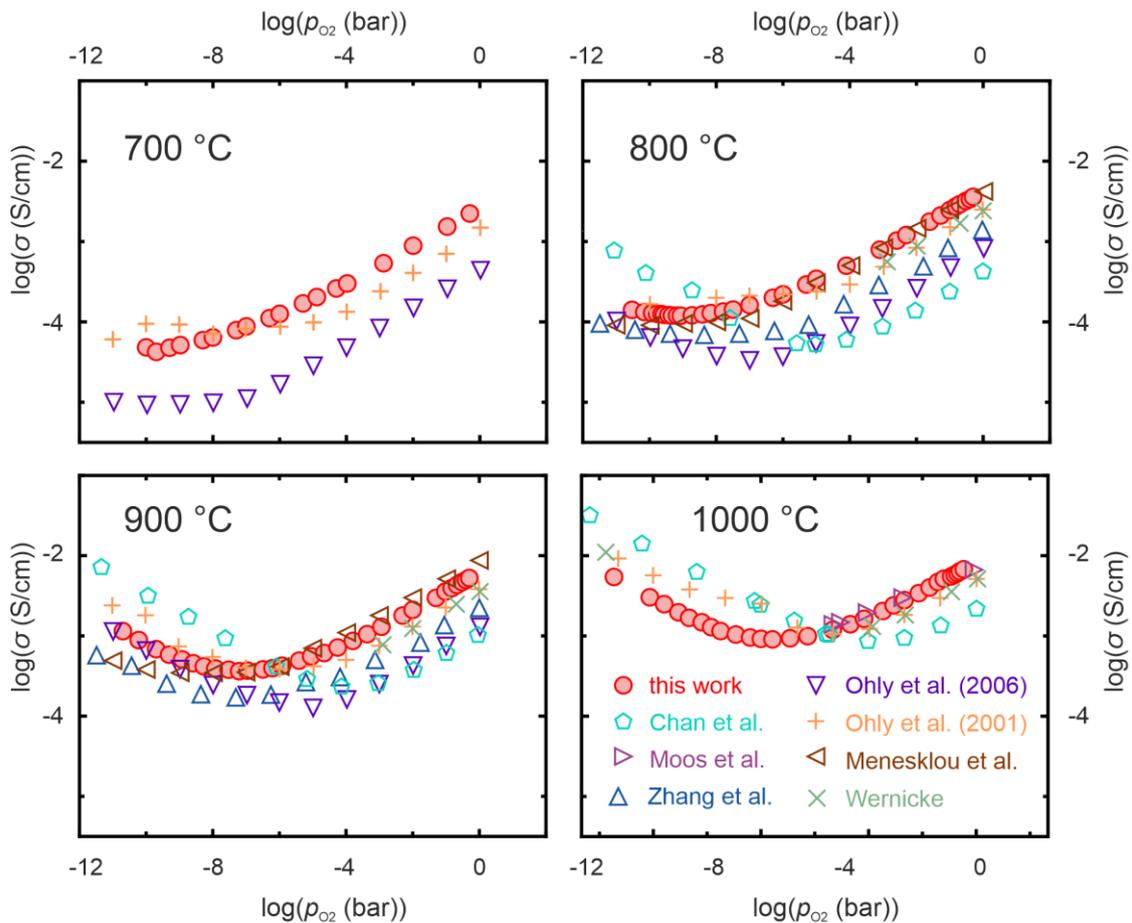

Figure S5. Comparison of the Brouwer diagrams of the conductivity of $SrTiO_3$ single crystals obtained in this work (●) with literature data. The compilation comprises data from undoped $SrTiO_3$ single crystals (▽) by Ohly et al. [3] and (+) [4] and by (×) Wernicke [5], from undoped polycrystalline $SrTiO_3$ (⬠) by Chan et al. [6], from undoped $SrTiO_3$ ceramics (▷) by Moos et al. [1], from polycrystalline $SrTiO_3$ doped with 0.01 at% Fe (◁) by Meneskou et al. [7] and from $SrTiO_3$ single crystals doped with 0.01 wt% Fe (△) by Zhang et al. [8].

A further proof that the applied measurement technique is capable of investigating the conductivity of the sample without changing its properties is the reproducibility of the data points. We measured the resistance of the sample several times in the selected pressure regimes, as indicated by the different symbols in Figure S4a. The data were obtained while either increasing or decreasing the pressure. The result shows that the data was highly reproducible, with a negligible deviation across the different measurement runs

V. Comparison of the HTEC results with the literature data

In order to demonstrate the compatibility of the physical method using pure oxygen dosed into a UHV chamber with conventional methods employing the use of gas mixtures, we compare our results of the equilibrium conductivity with data from the literature. In Figure S5, the Brouwer diagrams obtained at the temperatures of 700 °C, 800 °C, 900 °C and 1000 °C are shown, together with selected results from the literature. As only sparse data about the defect chemistry measurements on undoped $SrTiO_3$ is available, we also included data from ceramics and slightly acceptor-doped $SrTiO_3$. The comparison shows that the measured conductivity of all measurements lies in the same regime and the dependency on the oxygen partial pressure follows a concurrent trend. However, it can also be seen that there is a significant scatter in the data concerning the magnitude of the conductivity, as well as the slopes of the pressure-dependent conductivity curves and the position of the intrinsic minimum. Even when only considering the data for undoped single crystals, a significant variation between the different measurements can be identified, indicating that the defect chemistry of $SrTiO_3$ is very sensitive on external parameters such as crystal quality and preparation.

VI. Investigation of surface reactions

Our approach to investigating the sample inside a vacuum chamber allows for the direct correlation of the conductivity data with the investigations of, *e.g.*, the electronic structure and chemical composition measured under UHV conditions. Since the Brouwer diagram in Figure S4a showed the reversibility relative to the direction of exposure to the oxygen partial pressure (reduction or oxidation), the state of the surface layer at the starting point of the diagram at low pressure ($p_{O2} < 10^{-11}$ mbar) can be easily analyzed using surface-sensitive methods. After characterizing the $SrTiO_3$ (100) surface layer during thermal reduction under UHV conditions at high temperatures (700 °C-1000 °C), the following effects were found:

- The surface of an as received $SrTiO_3$ (100) crystal undergoes a permanent change in the ratio between SrO and $TiO_2$ and becomes Ti-rich with the increase in the annealing temperature [3]. As a result, the reconstruction of the surfaces changes from (1 × 1) to (√13 × √13) and ultimately to (√5 × √5).
- The initial presence of SrO monolayers on top of the crystal leads to a very strong chemisorption of $CO_2$, which results in the formation of $SrCO_3$ compounds [4]. During annealing under vacuum conditions, the $SrCO_3$ is again dissociated into SrO and $CO_2$ at temperatures between 800 °C and 900 °C. This means that in the investigations of the point defect behavior by HTEC experiments below 900 °C, we must accept the presence of complex boundary conditions for the interaction of oxygen and oxygen vacancies on $SrCO_3$ surfaces.
- The electronic structure of the surface layer changes strongly with ongoing vacuum reduction. As identified by XPS, with an information depth of approximately 6 nm, part of the Ti ions change their valence from +4 to +3 after slight reduction (Figure S6a) and eventually to +2 after strong reduction (Figure S6c). In both cases, the valence returns to +4 after exposure to oxygen at high temperature. This shows that the oxygen vacancies, which were generated upon reduction, are refilled and the electronic charge is transferred back from the Ti ion to the oxygen ion. At the same time, the electronic structure at the valence band also changes (Figure S6b, d). After a strong reduction, even the occupation of a continuum of $d^{1-2}$ states in the band gap can be detected (Figure S6f), reflecting metallic behavior [8]. After reoxidation, a surface bandgap in the range of 2.7 eV was found, which is smaller than that of the bulk (3.25 eV [9]). The special role of the surface layer also becomes obvious when comparing spectra obtained at different take-off angles (Figure S6e). It can be seen that the measurement at the angle of 22.5 °, which is more surface-sensitive, detects slightly more occupied $Ti^{+3}$ states than that at 45 °, indicating that there is a gradient in the surface layer induced by the reduction. However, the difference between the two angles is relatively small, which suggests that even in deeper parts of the surface below the information depth of 6 nm (typical for XPS using $Al_{K\alpha}$), a layer with the enrichment of $Ti^{+3}$ and $Ti^{+4}$ exists.

- The surface layer plays a decisive role in electronic transport in vacuum-reduced $SrTiO_3$ crystals [5, 6]. This effect relates to the easy reducibility of dislocations, which serve as conducting channels inside an insulating matrix. Due to inhomogeneities in the distribution of dislocations that are introduced during crystal growth and surface preparation by cutting and polishing, significant local variations in surface conductivity and surface potential can be represented at the nanoscale [3, 7]. Hence, in some surface areas, the exchange reactions between the crystal and surrounding oxygen atmosphere are more favorable than in others.

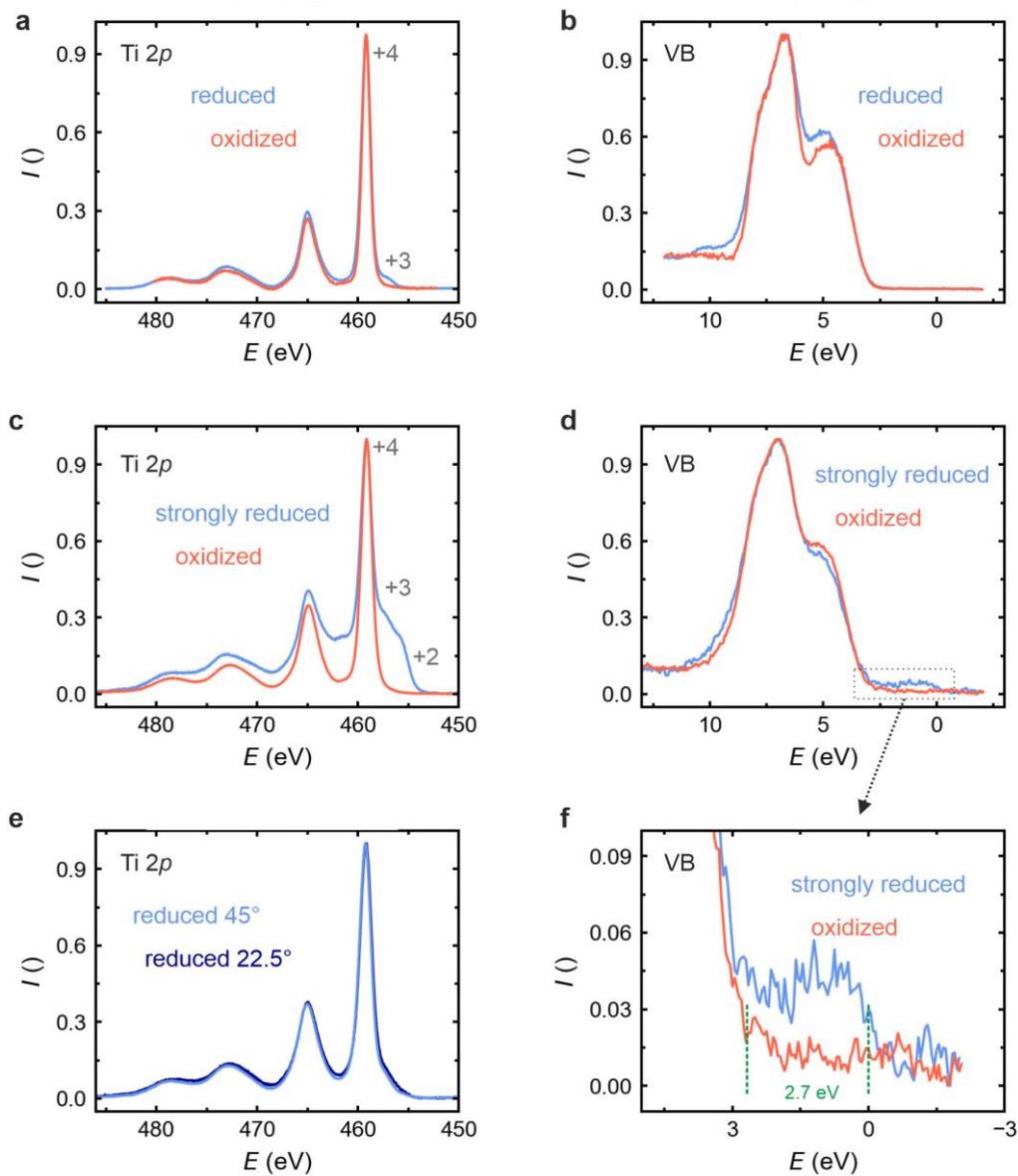

Figure S6. Analysis of the electronic structure of the surface by XPS. Normalized Ti2$p$ core level and valence band spectra after thermal reduction and reoxidation at 1000 °C (a-b) and after long-term thermal reduction and reoxidation at 1000 °C (c-d), with a magnification of the band gap region (f). Comparison between the Ti2$p$ core level spectra of a reduced crystal measured at different take-off angles (e).

## VII. Extraction of electrical conductivity relaxation traces

As we continuously measured the conductivity of the sample while manually changing the $O_2$ pressure inside the chamber, additional data processing was necessary for further analysis. In particular, it was necessary to recalibrate the starting time for each pressure step. The adjustment of the pressure inside the chamber by opening the valve to the oxygen supply and simultaneously adjusting the valve between the chamber and vacuum pump by hand led to an unavoidable delay of several seconds until the pressure in the chamber was equilibrated. For further research, a significant improvement in the data quality of the conductivity relaxation traces can be expected if the pressure is automatically adjusted using a dedicated computer-controlled valve system. In order to evaluate the manually-obtained data of the present study, we used the tangent method to determine the starting time of the pressure step. Figure S7a shows the measured bulk conductivity at 1000 °C for the pressure step between $1.0 \times 10^{-4}$ bar and $3.1 \times 10^{-4}$ bar. At first, the normalized macroscopic conductivity was calculated using Eq. 9 of the main manuscript, as is shown in Figure S7b. It can be seen that the normalized conductivity increased nearly linearly in the range of $0.2 < \bar{\sigma} < 0.5$. By linear regression, this range was fitted and thus the tangent of the data was obtained. We then defined starting point of the step for the fitting of the conductivity relaxation as the point where the tangent is zero.

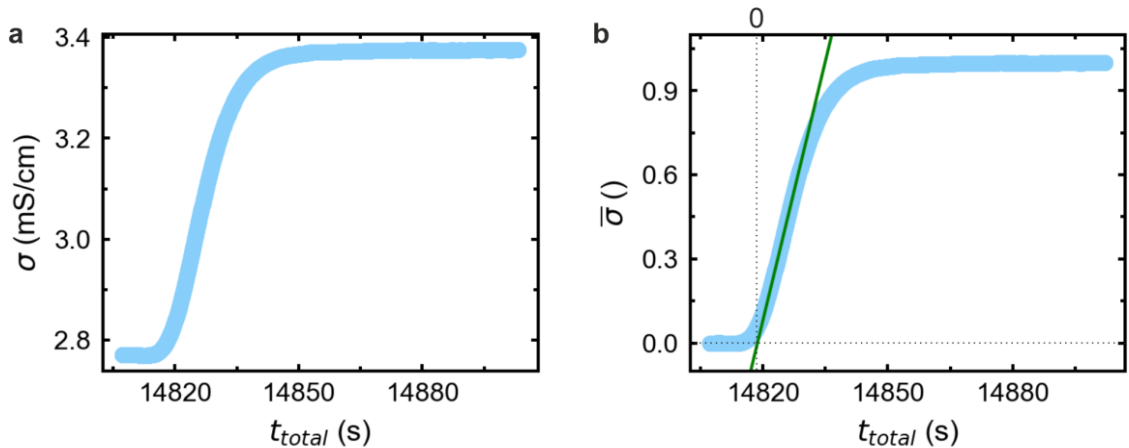

Figure S7. Estimation of the starting time of the oxygen pressure step using the tangent method. a) Measured conductivity; b) normalized conductivity and linear fit of the tangent (green line). The step from $1.0 \times 10^{-4}$ bar to $3.1 \times 10^{-4}$ bar at 1000 °C is also shown.

VIII.    Simulation of the electrical conductivity relaxation traces

The electrical conductivity traces obtained for each pressure step were simulated using the equation for the change in the normalized conductivity upon oxidation as derived by Yasuda et al [10]

$$\bar{\sigma}(t) = 1 - \sum_{i,j,k}^{\infty} \frac{2L_x^2 \exp\left(\frac{-\beta_i^2 Dt}{l_x^2}\right)}{\beta_i^2(\beta_i^2 + L_x^2 + L_x)} \cdot \frac{2L_y^2 \exp\left(\frac{-\beta_j^2 Dt}{l_y^2}\right)}{\beta_j^2(\beta_j^2 + L_y^2 + L_y)} \cdot \frac{2L_z^2 \exp\left(\frac{-\beta_k^2 Dt}{l_z^2}\right)}{\beta_k^2(\beta_k^2 + L_z^2 + L_z)} \quad (13)$$

with

$$L_x = \frac{l_x k}{D} = \beta_i(\tan \beta_i) \,, \; L_y = \frac{l_y k}{D} = \beta_j(\tan \beta_j) \,, \; L_z = \frac{l_z k}{D} = \beta_k(\tan \beta_k) \quad (14)$$

where $l_x$, $l_y$, $l_z$ are the sample dimensions, $k$ is the surface exchange coefficient and $D$ is the chemical diffusion coefficient of oxygen in the solid. The $\beta$ were calculated as the first ten non-zero solutions of Eq. 14 for each dimension. In this way, the traces were fitted using the two free parameters $k$ and $D$ and least squares fitting. The results of the simulation are shown in Figure S8. It can be seen that only the exchange coefficient $k$ can be determined within realistic error boundaries. As was already discussed in the main text, with the help of the squared error matrix, the simulation is not sensitive to the diffusion coefficient $D$, which indicates that the surface reaction determines the kinetics of the oxidation of the sample.
In Figure S9-Figure S12, the relaxation curves for the four different temperatures of 700 °C, 800 °C, 900 °C and 1000 °C are shown together with the fit curve. For most of the pressure steps, a reasonable agreement between simulation and experimental data is achieved. In cases where the absolute conductivity change is small, *e.g.*, for 800 °C below a pressure of $10^{-9}$ bar (*cf.* Figure S10), the fitting of the data becomes difficult. In some traces, *e.g.*, at 900 °C in a pressure regime of $10^{-6}$-$10^{-8}$ bar (*cf.* Figure S11), it can also be seen that the conductivity at first decreases from the initial value before it increases. This probably relates to the manual adjustment of the pressure in the chamber as described above, which could lead to an initial reduction of the oxygen pressure before the new equilibrium at higher pressure is achieved. This shows, as a proof of principle, that the measurement method is feasible to obtain electric conductivity relaxation traces, and that the quality of the data could be significantly improved in the future if a faster and more precise pressure control system could be implemented.

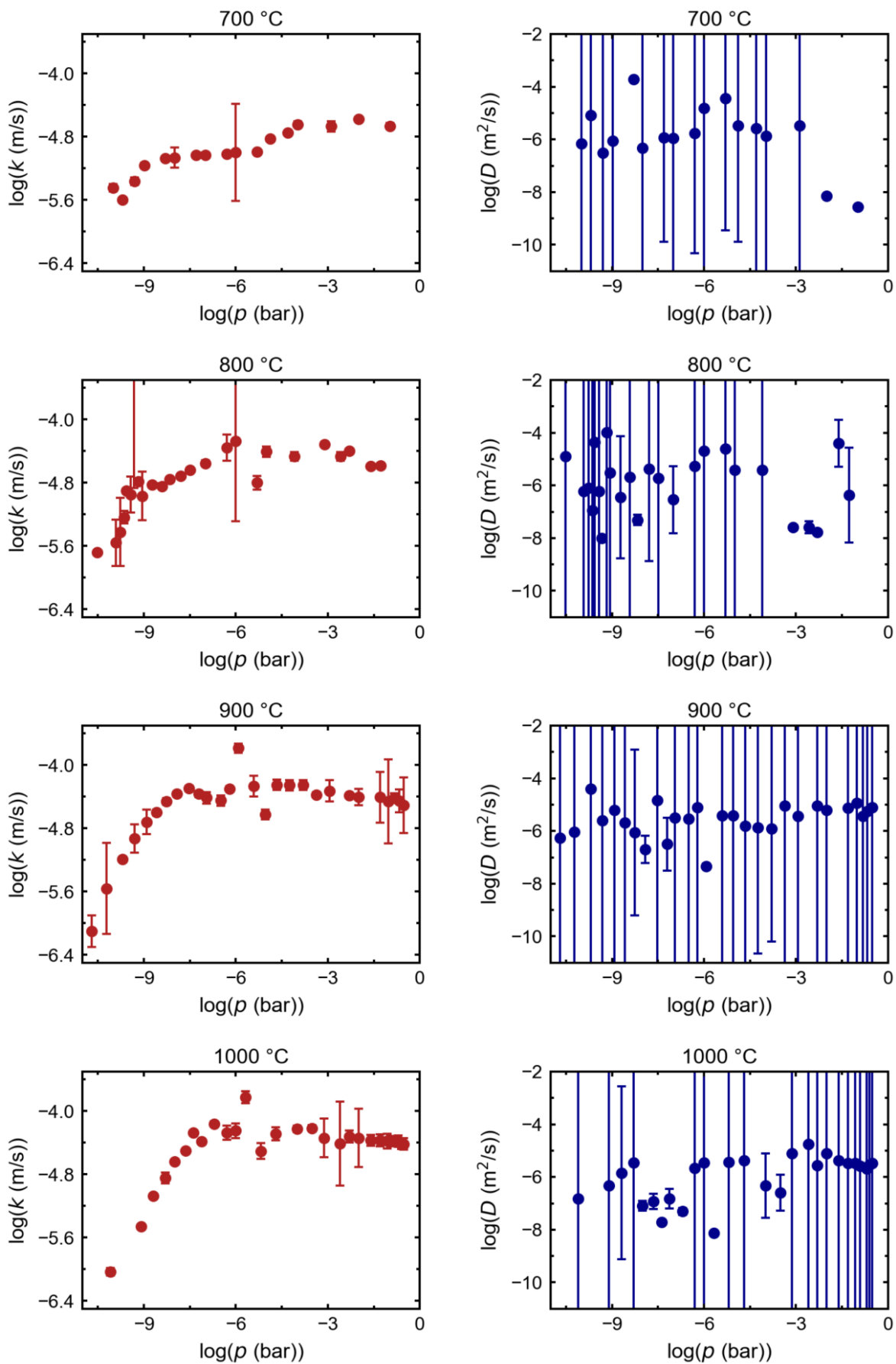

Figure S8. Simulation of the electric conductivity relaxation traces. The fit parameters $k$ and $D$ are shown as a function of oxygen pressure for different temperatures.

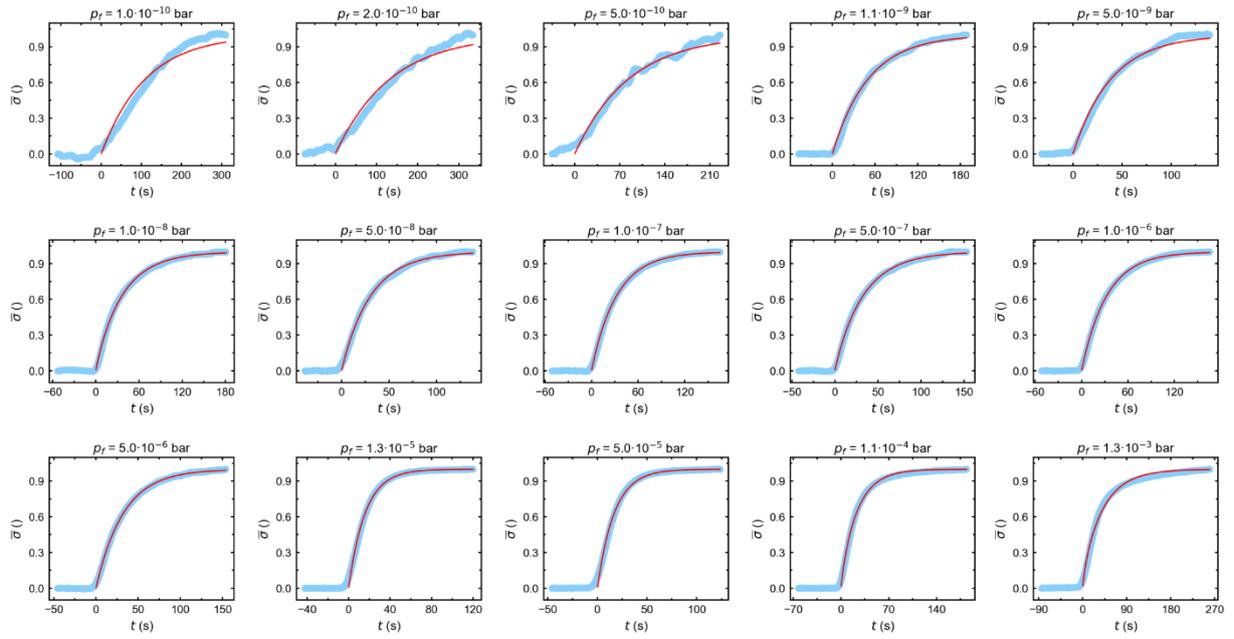

Figure S9. Electrical conductivity relaxation traces of the normalized conductivity (blue) and simulation result (red) for a temperature of 700 °C. The final pressure of each pressure step is denoted by $p_f$.

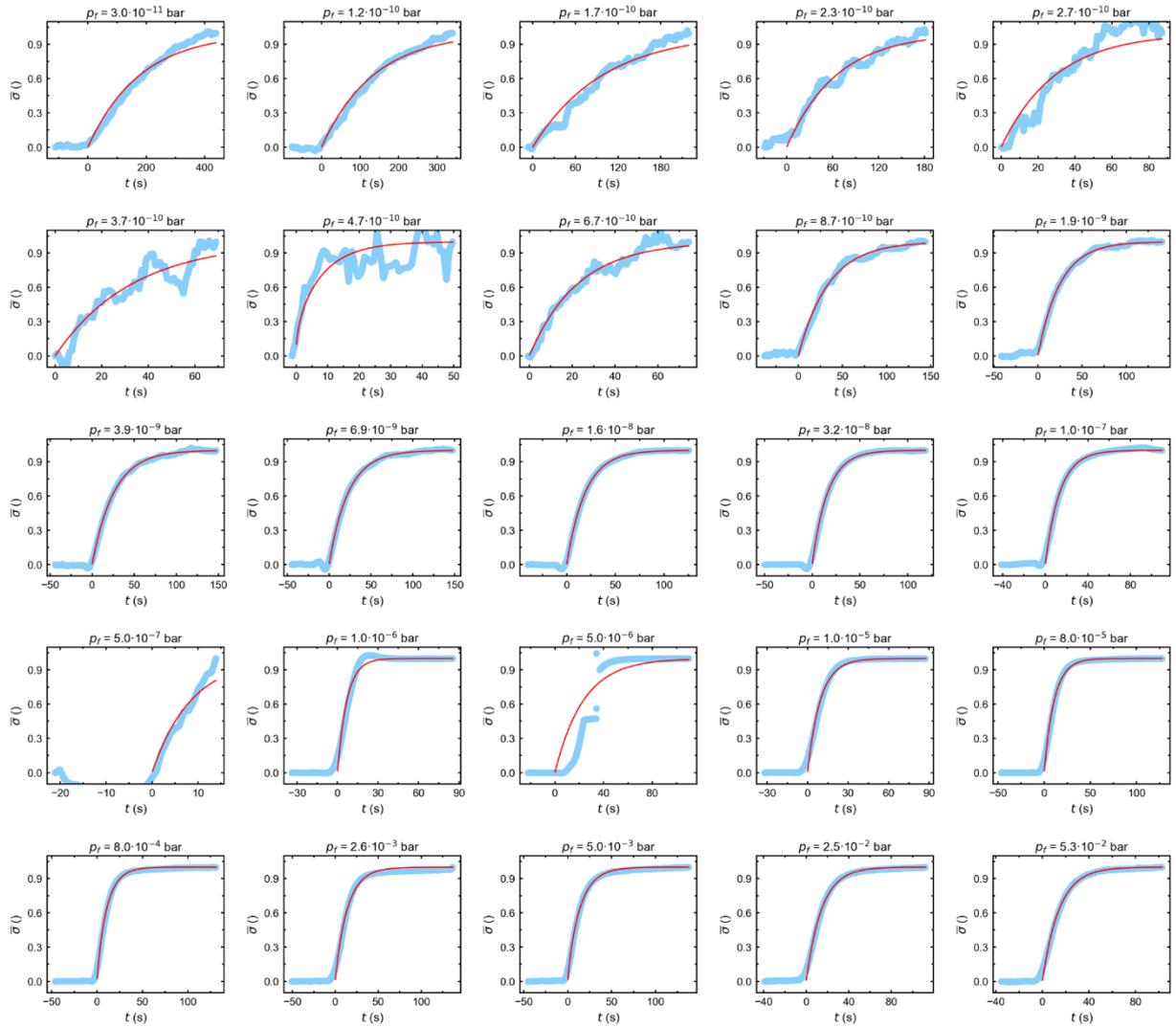

Figure S10. Electrical conductivity relaxation traces of the normalized conductivity (blue) and simulation result (red) for a temperature of 800 °C. The final pressure of each pressure step is denoted by $p_f$.

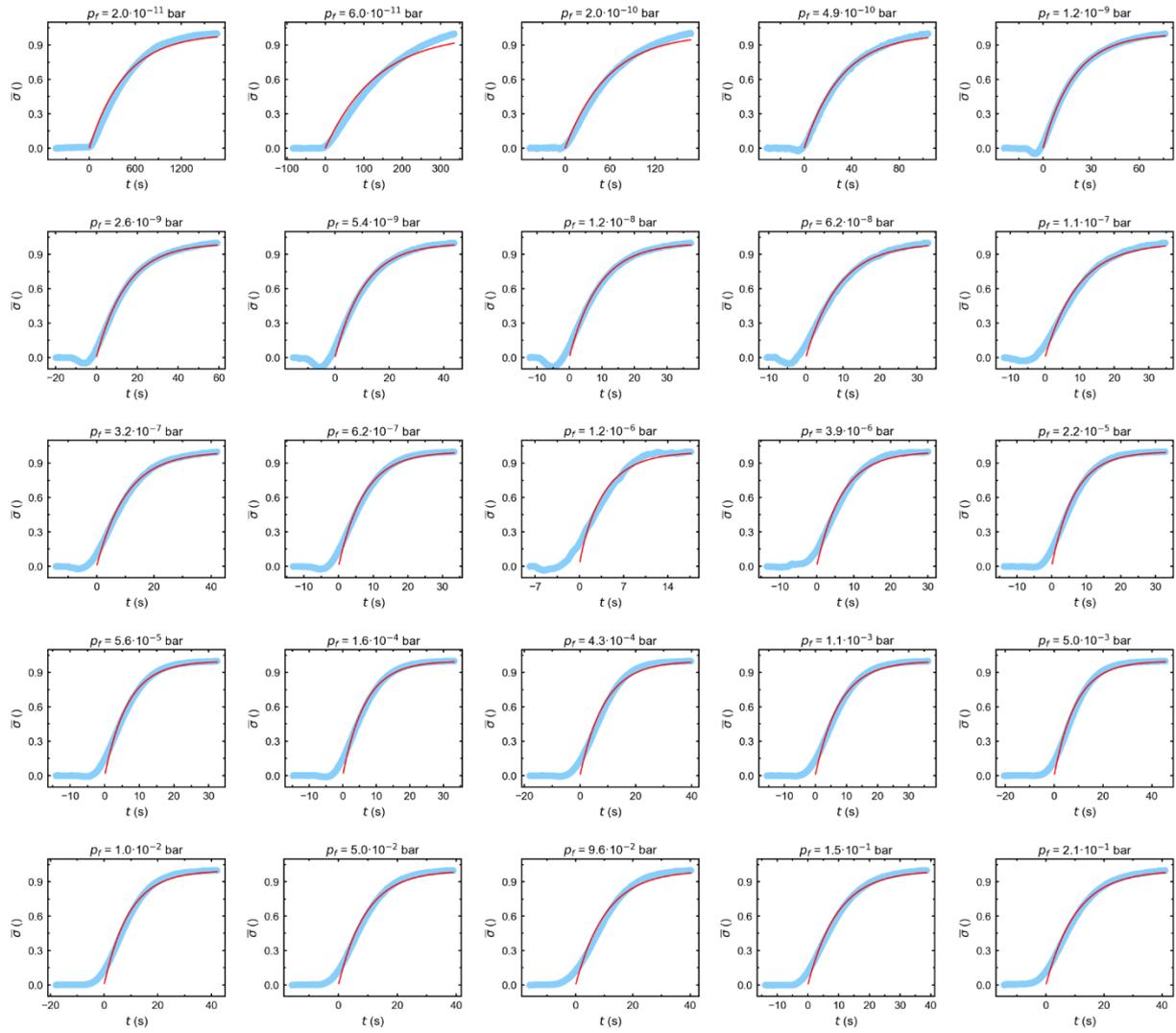

Figure S11. Electrical conductivity relaxation traces of the normalized conductivity (blue) and simulation result (red) for a temperature of 900 °C. The final pressure of each pressure step is denoted by $p_f$.

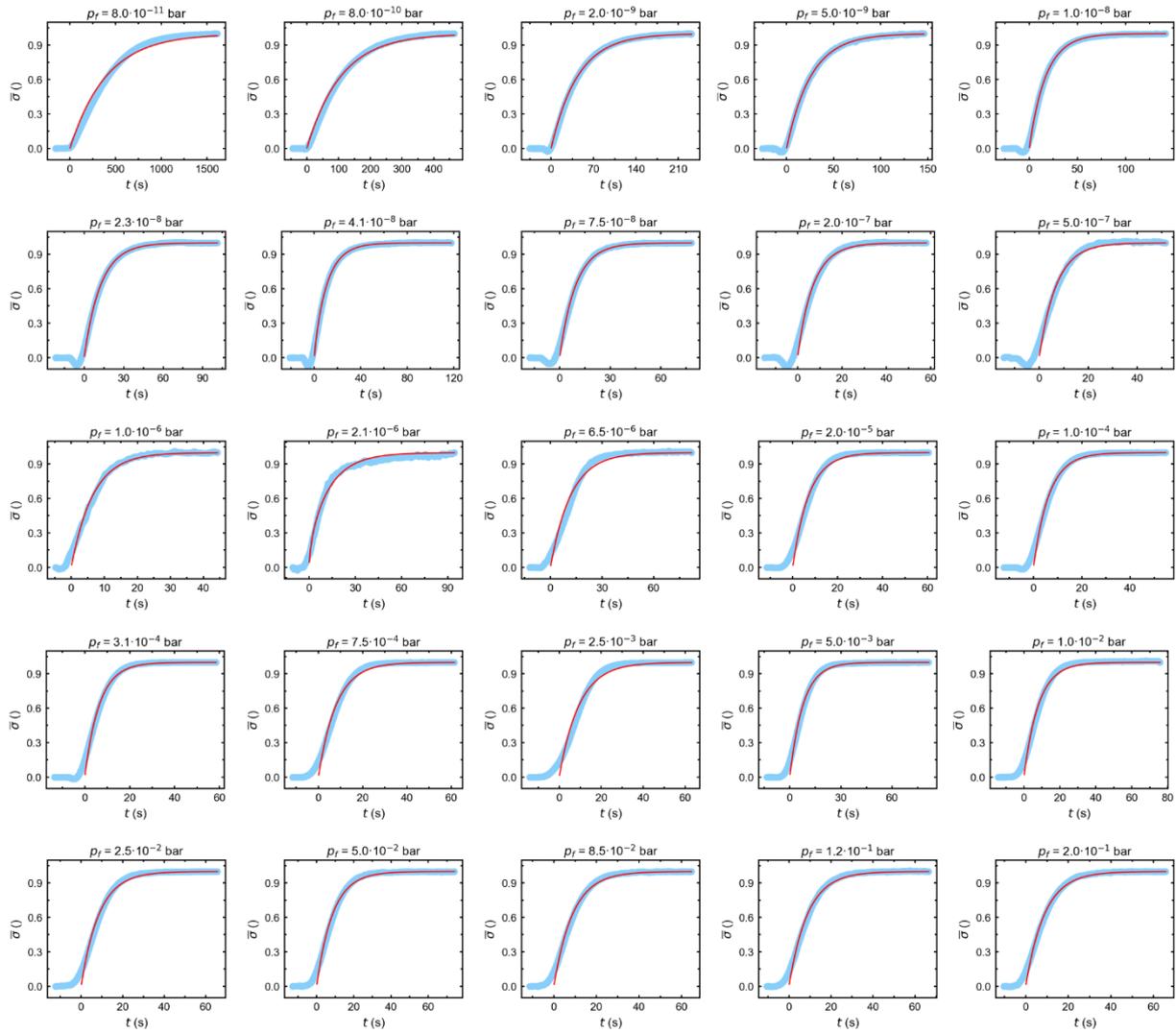

Figure S12. Electrical conductivity relaxation traces of the normalized conductivity (blue) and simulation result (red) for a temperature of 1000 °C. The final pressure of each pressure step is denoted by $p_f$.